\documentclass[12pt,preprint]{aastex}
\shorttitle{On the WD mass problem of CVs}
\shortauthors{Liu \& Li.}

\begin{document}

\title{On the White Dwarf Mass Problem of Cataclysmic Variables}
\author{Wei-Min Liu\altaffilmark{1,2,3} and Xiang-Dong Li\altaffilmark{1,3}}
\altaffiltext{1}{Department of Astronomy, Nanjing University,
Nanjing 210046, China} \altaffiltext{2}{Department of Physics, Shangqiu Normal
University, Shangqiu 476000, China} \altaffiltext{3}{Key Laboratory of of Modern
Astronomy and Astrophysics, Ministry of Education, Nanjing 210046,
China}

\begin{abstract}

Recent observations show that the white dwarfs (WDs) in cataclysmic Variables (CVs) have an average mass significantly higher than isolated WDs and WDs in post-common envelope binaries (PCEBs), which are thought to the progenitors of CVs. This suggests that either the WDs have grown in mass during the PCEB/CV evolution or the binaries with low-mass WDs are unable to evolve to be CVs. In this paper, we calculate the evolution of accreting WD binaries with updated hydrogen accumulation efficiency and angular momentum loss prescriptions. We show that thermal timescale mass transfer is not effective in changing the average WD mass distribution. The WD mass discrepancy is most likely related to unstable mass transfer in WD binaries in which an efficient mechanism of angular momentum loss is required.

\end{abstract}


\section{Introduction}

Cataclysmic variables (CVs) are compact binaries containing an accreting white dwarf (WD) and a low-mass main-sequence (MS) donor star \citep[see][for reviews]{war95,rit10,kni11}. Since the maximum size that the progenitor of the WD can reach during its evolution is much larger than the current binary separation, it is
believed that the binary orbit must have significantly shrunk through so-called common envelope (CE) evolution, during which the primary star has lost most of its envelope and left a degenerate core, i.e., the proto-WD \citep{pac76}. If surviving CE evolution, the post-common envelope binary (PCEB) can evolve to be a CV by two ways. (1) If the secondary is originally of low-mass ($\lesssim 1.1\,M_{\sun}$), angular momentum loss (AML) via magnetic braking (MB) and/or gravitational radiation (GR) may reduce the orbit so that the secondary overflows its Roche-lobe (RL) and initiates mass transfer within a Hubble time. (2) A more massive secondary star can fill its RL due to nuclear expansion. In that case the mass transfer first proceeds on a thermal timescale of the donor star, and the binary may appear as a super-soft X-ray source \citep[SSS,][for a review]{Kah1997}. After the mass ratio reverses, the binary evolution may follow the similar path as in the former case.

The WD mass problem concerns the discrepancy in the mass distribution for isolated WDs and WDs in CVs, which is in contradiction with theoretical expectations \citep{de92,pol96}. While the average mass of isolated WDs is $\sim0.6~M_\odot$ \citep[e.g.,][]{ber92,kep07}, the WDs in CVs are remarkably more massive, with an average mass $\sim 0.8\,M_\odot$ \citep[][and references therein]{kni11,zor11}, and this discrepancy is unlikely to be due to the observational bias toward CVs with more massive WDs, which tend to be brighter and hence easier to be discovered \citep{zor11}.

Considering the fact that WDs in PCEBs have an average mass around $0.67\,M_\sun$, similar to that of isolated WDs \citep{zor11}, there are three possible resolutions to the WD mass problem. (1) The WDs have increased their masses during the CV evolution, but this idea is disfavored since numerical calculations \citep[e.g.,][]{pra95,yar05} have shown that all (or even more) of the accreted matter is lost during nova outbursts. (2) A significant fraction of CVs originate from PCEBs with initially more massive secondaries \citep{sch02}, and during the thermal-timescale mass transfer phase there is stable burning of hydrogen and helium on the surface of the WDs, leading to mass growth. Using a binary population synthesis (BPS) method, \cite{wij15} found that this  model can produce a large number of massive WDs, but still with too many helium WDs and evolved donor stars, both inconsistent with observations. (3) Mass transfer in PCEBs with low-mass WDs are dynamically unstable, provided that there is extra AML besides MB and GR. Such binaries would merge rather than evolve into CVs \citep{sch16,nel16}.

The aim of this paper is to examine the feasibility of the above models with detailed binary evolution calculations. For resolution (2) we note that \citet{wij15} employed the semi-analytic method of \citet{hur02} \citep[with modifications by][]{cla14} to deal with thermal-timescale mass transfer. The actual evolution of the mass transfer rate should be more complicated than this simplified estimate \citep[e.g.,][]{xu07}. Moreover, a semi-analytic criterion for stable hydrogen burning \citep{hac99,meng09} was adopted, which possibly overestimates the net accumulation efficiency. Recently, \citet{hil16} presented the results of the parameter space for net mass accumulation through simulations of a long series of hydrogen and helium flashes on WDs with mass $\sim 0.6-1.4\,M_\sun$. We employ them to calculate how much mass a $0.6-0.7\,M_\sun$ WD in binaries can grow. For resolution (3), \citet{sch16} showed that assuming an empirical model for consequential angular momentum loss (CAML) that increases with decreasing WD mass can greatly reduce the number of CVs with low-mass WDs. If this idea is correct, the effect of the CAML should also be present in CVs. \citet{kni11} recently also reconstructed the CV evolution by use of the mass-radius relation for the secondary stars, which can be used to constrain the CAML model parameters. The result is also helpful in understanding the physics behind the CAML.

This paper is organized as follows. We describe the input physics included in the binary evolution calculations in Section 2. The calculated results on the WD mass growth are presented in Section 3. In Section 4, we examine the parameter space for dynamically unstable mass transfer under CAML and its influence on the CV evolution. Our conclusions and discussion are in Section 5.

\section{Input physics}

We use an updated stellar evolutionary code developed by \cite{Egg71,Egg72,Egg73}
to follow the evolution of PCEBs, which consist of a CO WD of mass
$M_{\rm WD}$ and a  MS donor star of mass $M_2$. We assume solar chemical composition
($X = 0.70$, $Y = 0.28$, and $Z = 0.02$) for the donor star. The ratio of the mixing length to the pressure
scale height and the convective overshooting parameter are set to be 2.0 and 0.12, respectively. The stellar
OPAL opacities are taken from  \cite{rog92} and \cite{ale94}.

The growth of an accreting WD depends on how much mass can be accumulated during the hydrogen and helium burning phases, and there have been extensive investigations on this subject \citep{kp94,pra95,yar05,st12,id13,pi14,hil15,hil16}. The mass growth rate $\dot{M}_{\rm WD}$  of a WD can be described as follows,
\begin{equation}
\dot{M}_{\rm WD}=\eta_{\rm H}\eta_{\rm He}|{\dot M}_2|,
\end{equation}
where $\eta_{\rm H}$ is the accumulation efficiency during hydrogen burning, $\eta_{\rm He}$  the accumulation efficiency during helium burning, and ${\dot M}_2$ the mass transfer rate.

\cite{hil16} simulated the long-term evolution of WDs with a range of masses ($0.65-1.4M_\sun$) and accretion rates ($3\times 10^{-8}-6\times 10^{-7}\,M_\sun$\,yr$^{-1}$) of hydrogen-rich material to determine the efficiency of mass retention for each parameter combination.  We fit their numerically calculated results for $\eta_{\rm H}$. If the mass accretion rate exceeds $6\times 10^{-7}\,M_\sun$\,yr$^{-1}$, we fix the hydrogen accumulation efficiency to be that for $6\times 10^{-7}\,M_\sun$\,yr$^{-1}$, as this rate is actually higher than  the traditional critical accretion rate for stable hydrogen burning on a $<1M_\sun$ WD considered here \citep[e.g.,][]{n07}. We adopt the prescriptions suggested by \cite{Ka2004} for $\eta_{\rm He}$, because the calculations of \cite{hil16} for helium flashes were only limited to $1\,M_\sun$ and $1.2\,M_\sun$ WDs. A comparison of the magnitudes of $\eta_{\rm He}$ with the same $M_{\rm WD}$ and $\dot{M}$ in \cite{Ka2004} and \cite{hil16} shows that the former is generally higher, so our following results on the WD growth might be regarded as upper limits.

In our calculation, the excess matter $(1-\eta_{\rm H}\eta_{\rm He})|{\dot M}_2|$ is assumed to be ejected from the WD in the form of isotropic wind, carrying its specific AM \citep{hac96,sob97}. AML due to GR \citep{ll75} and MB \citep{verb81} is also included.

\section{Calculated results}

In our calculations, we set the initial WD masses to be $0.6 - 0.7 M_\odot$ to examine its evolution with mass accretion and thermonuclear burning. This mass range  is roughly in accord with the average mass of WDs in PCEBs. We do not consider lower-mass WDs  since reasonable estimate of the accumulation efficiencies are not available for them. The initial masses of the secondary/donor stars are chosen to be in the range of $0.3 - 2.5M_\odot$. Figures 1 summarizes the outcome of the binary evolution in the initial orbital period - donor mass plane for a $0.6 M_\odot$  WD. The parameter space is divided into five regions: (1) the forbidden region where the secondary's radius exceeds its RL radius at the beginning, (2) the region where the mass transfer is subject to (delayed) dynamical instability, (3) the region with no RL-overflow within a Hubble time, (4) the region with the  WD mass growth $<0.01 M_\sun$, and (5) the region where the WD can effectively increase mass (by more than $0.01 M_\sun$). The red and blue circles represent the final WD masses in the range of $0.6-0.8 M_\sun$ and $0.8-1.0 M_\sun$, respectively. The red and blue solid triangle denote the binaries with same mass growth range respectively, but unable to evolve into CVs.
Figure~2 is similar to Fig.~1 but for $0.7 M_\odot$  WDs. The red, blue, and pink circles represent the final WD masses in the range of $0.7-0.9 M_\sun$, $0.9-1.1 M_\sun$, and $1.1-1.2 M_\sun$, respectively. The triangles have the same meanings as in Fig.~1.

Details of the evolutionary sequences for different sets of initial parameters are demonstrated in Figs.~3-7. In each figure, the left and right panels exhibit the evolution of the mass transfer rate (solid line) and the white dwarf mass (dashed line), and the evolution of the donor mass (solid line) and the orbital period (dashed line), respectively. In Fig.~3, the binary initially consists of a $0.7 M_\sun$ WD and a $0.6 M_\sun$ secondary in a 1.26 day orbit. Mass transfer is driven by MB at a rate $<10^{-8}\,M_\sun$ yr$^{-1}$, and the WD has no chance to grow mass, so only part of the evolutionary tracks are plotted here. In Figs.~4-6 the donor mass is increased to be $1.2M_\sun$, $1.7M_\sun$, and $2.0M_\sun$, respectively. They represent the cases of the WD growth with increasing amount. Their distinct feature is the early SSS stage when the donor star is more massive than the WD. Larger mass ratio $q=M_2/M_{\rm WD}$  leads to higher mass retention efficiency,  and the final WD masses in these binaries are $0.715 M_\odot$, $0.819 M_\odot$, and $1.119 M_\odot$, respectively. In the subsequent phases the mass transfer rates decreases continuously and the WD masses do not change.  In Fig.~7 the donor mass is taken to be $2.3 M_\sun$. The mass transfer firstly proceeds on a thermal timescale, then rises rapidly due to delayed dynamical instability.

To roughly estimate the fraction of PCEBs with efficient WD mass growth by thermal-timescale mass transfer, we incorporate our calculated results with the BPS study of PCEBs and CVs by \cite{pol07}. According to Figs.~2 and 3 in that paper, we find that with  the CE efficiency parameter
$\alpha_{\rm CE}$=0.2 and 0.6, about $(17-24)\%$ of the binaries with an initial $0.6 M_\odot$ WD can evolve into CVs with the final WD mass $>0.7M_\odot$. For an initial $0.7 M_\odot$ WD, around $(8-9)\%$ of the binaries can evolve into CVs with the final WD mass $\geq0.8M_\odot$. Thus thermal-timescale mass transfer seems unable to significantly change the distribution of the WD mass.

\section{Unstable mass transfer in PCEBs}

We then move to resolution (3). In the standard model of the CV evolution, AML is caused dominantly by MB above the period gap and solely by GR below the period gap \citep{Rappaport1983,spr83}. Extra AML might amplify the mass transfer rate or even cause the mass transfer to be dynamically unstable, since the RL radius of the donor star shrinks more rapidly than the stellar radius \citep{sob97}. \cite{sch16} proposed that, for a specific form of CAML that increases with decreasing WD mass, the parameter space for (delayed) unstable mass transfer can be significantly enlarged, especially for PCEBs with low-mass WDs. Thus these binaries are likely to merge after experiencing a rapid mass transfer phase. This increases the average mass of the WDs in binaries that successfully evolve to be long-lived CVs.

By following the evolutionary paths of CVs, \citet{kni11} found the the best fit of the observational data gives the AML rates to be $2.47(\pm0.22)$ times the GR rate below the period gap and $0.66(\pm0.05)$ times the MB rate above the period gap, indicating that extra AML mechanism(s) works at least below the period gap if we believe that the MB effect has switched off.
This kind of AML is most likely to be related to mass loss during the CV evolution. Assuming that all the transferred mass is lost from the binary during nova outbursts taking the specific AM of the WD, we can write the mass transfer rate as follows \citep{Rappaport1983}
\begin{equation}
-\frac{\dot{M}_{2}}{M_{2}}=\frac{\frac{1}{2}\left(\frac{\dot{R_{2}}}{R_{2}}\right)_{\rm ev,th}
-\left(\frac{\dot{J}_{\rm sys}}{J}\right)}{\frac{5}{6}+\frac{\zeta}{2}-\frac{q}{3(1+q)}-\frac{q^2}{1+q}},
\end{equation}
where $(\dot{R_{2}}/R_{2})_{\rm evol,th}$ represents the change in the donor star's radius due to thermal or nuclear evolution, $J$ is the binary orbital AM,
$\dot{J}_{\rm sys}$ the systemic AML rate caused by MB and/or GR, and $\zeta$ the adiabatic mass-radius exponent of the donor star (i.e., $R_2\sim M_2^{\zeta}$). For low-mass MS stars, we can neglect the change in the stellar radius due to evolution, i.e., $(\frac{\dot{R}_{2}}{R_2})_{\rm ev,th}=0$. For CVs below the period gap, let $\dot{J}_{\rm sys}=2.47\dot{J}_{\rm GR}$ (where $\dot{J}_{\rm GR}$ is the rate of AML by GR) we can rewrite Eq.~(2) as
\begin{equation}
\frac{\dot{M}_{2}}{M_{2}}=\frac{\left(\frac{2.47\dot{J}_{\rm GR}}{J}\right)}{\frac{5}{6}+\frac{\zeta}{2}-\frac{q}{3(1+q)}-\frac{q^2}{1+q}}.
\end{equation}
However, since the normalized GR-induced AML rate should not be larger than unity, this phenomenological form simply reflects some more complicated processes that drive the mass transfer. \citet{shao12} examined several potential CAML mechanisms for mass loss including isotropic wind from the WD, outflow from (inner/outer) Lagrangian points, and formation of a circumbinary (CB) disk. They showed that formation of a CB disk or outflow from the outer Lagrangian point seems to be able to account for the extra 1.47GR AML rate. Note that \citet{shao12} considered each CAML mechanism separately, while in real situation there may be different ways of mass loss simultaneously \citep[e.g.,][]{wil13}.
In the following we assume that mass loss occurs in two ways: (1) a fraction $\delta$ of the lost mass leaves the binary in the form of isotropic wind from the WD, and (2) the other $(1-\delta)$ part forms a CB disk surrounding the system\footnote{Mass loss from the outer Lagrangian point has the similar effect.}. The total rate of the CAML is then given by
\begin{equation}
\frac{\dot{J}_{\rm CAML}}{J}=\left[\frac{\delta q^2}{1+q}+\gamma(1-\delta)(1+q)\right]\frac{\dot{M}_2}{M_2},
\end{equation}
where $\gamma$ = 1.5 is the ratio of the radius of the CB disk to the binary separation \citep{sob97}. The first and second terms on the right hand side of Eq.~(4) represent the contribution from isotropic wind and the CB disk, respectively. Similar to Eq.~(2), we can write the mass transfer rate for CVs below the period gap as
\begin{equation}
\frac{\dot{M}_{2}}{M_{2}}=\frac{\left(\frac{\dot{J}_{\rm GR}}{J}\right)}{\frac{5}{6}+\frac{\zeta}{2}-\frac{q}{3(1+q)}-\left[\frac{\delta q^2}{1+q}+\gamma(1-\delta)(1+q)\right]}.
\end{equation}
Since Eq.~(3) and (5) should match each other, we get
\begin{equation}
2.47\left\{\frac{5}{6}+\frac{\zeta}{2}-\frac{q}{3(1+q)}-\left[\frac{\delta q^2}{1+q}+\gamma(1-\delta)(1+q)\right]\right\}
=\frac{5}{6}+\frac{\zeta}{2}-\frac{q}{3(1+q)}-\frac{q^2}{1+q}.
\end{equation}
After transformation, we obtain
\begin{equation}
\delta=\frac{\gamma(1+q)^2-F+[F(1+q)-q^2]/2.47}{\gamma(1+q)^2-q^2},
\end{equation}
where $F=\frac{5}{6}+\frac{\zeta}{2}-\frac{q}{3(1+q)}$.

Figure 8 shows $1-\delta$, the fraction  of the lost matter that forms a CB disk,  as a function the mass ratio $q$ for $M_{\rm WD}$ = 0.6 $M_\odot$ and $\zeta\simeq 0.6$. It can be seen that $1-\delta$ decreases with $q$, ranging from about 0.1 to 0.4 when $q<0.5$. We then calculate the mass transfer sequences in a number of CVs adopting typical values of $\delta$. We regard the mass transfer to be dynamically unstable if the mass transfer rate rapidly rises to $>10^{-4}\,M_\sun$yr$^{-1}$. In Fig.~9 we show the boundary between stable and unstable mass transfer in the $M_2-q$ plane. The thin solid and dotted lines correspond to $1-\delta=0.2$ and 0.3, respectively. Also shown with the thick solid line is the stability limit in \citet{pol96}.  Compared with \citet{pol96}, it can be seen that for low-mass donors, the critical mass ratio for dynamically unstable mass transfer becomes significantly smaller when mass loss through a CB disk is considered. This can be understood as follows. It is evident that if the denominator of the right hand of Eq.~(5) is $<0$ then the mass transfer is dynamically unstable \citep{Rappaport1983}. This means that the mass transfer is stable only when
\begin{equation}
\delta>\frac{\gamma(1+q)^2-(1+q)F}{\gamma(1+q)^2-q^2},
\end{equation}
or
\begin{equation}
1-\delta<\frac{(1+q)F-q^2}{\gamma(1+q)^2-q^2}.
\end{equation}
When $q=0.1$, 0.5, and 1, we obtain $1-\delta<0.67$, 0.41, and 0.19, respectively. Thus PCEBs possessing low-mass WDs, which tend to have higher initial $q$ \citep{zor11}, are more likely to undergo unstable mass transfer for a given $1-\delta$.  Figure 10 shows an example of the evolutionary sequence with runaway mass transfer for a binary with $M_{\rm WD}=0.5,M_\sun$, $M_2=0.3,M_\sun$, $P_{\rm orb}=0.28$ day, and $1-\delta=0.2$.

\section{Discussion}

In this work we examine the proposed explanations for the WD problem in CVs by calculating the pre-CV evolution with updated hydrogen accumulation efficiency and AML prescription. Simulations by \citet{hil16} on hydrogen flashes disfavor the possibility of mass growth during the CV evolution, since the mass transfer rates  are too low to allow stable burning on WDs. However, we caution that there exist a group of SSSs (e.g., RX J0439.8$-$6809, 1E0035.4$-$7230, RX J0537.7$-$7034) whose properties are similar to those of ordinary CVs, suggesting that the mass transfer rates in these systems are (at least temporarily) high enough for stable hydrogen burning \citep[][and references therein]{abl14}. A possible mechanism is there is excited wind from the donor star driven by X-ray irradiation from the accreting WD \citep{tk98}. However, this self-excited wind requires some special conditions and should not be popular in CVs \citep{kt98}.

Our calculations demonstrate that in a minor fraction of PCEBs the WDs can effectively grow in mass through thermal-timescale mass transfer. This means that the formation channel from binaries with initially more massive secondaries may be less efficient than in \citet{wij15}. The main reason for the discrepancy is that the hydrogen accumulation efficiency of \citet{hil16} is considerably lower than that used by \citet{wij15}.

The best-fit AML prescriptions for present CVs by \citet{kni11} suggest that there should be extra AML mechanism, which is most likely related to mass loss during the binary evolution.
\citet{sch16} proposed a parameterized CAML with the specific AML inversely proportional to the WD mass, and showed that the criterion for stable mass transfer can be significantly altered compared with in the traditional model. The consequence is that a large fraction of the binaries with low-mass WDs would undergo unstable mass transfer and probably merge with their companion stars.

The physical mechanism to account for this extra AML is still open, and there are several possibilities proposed in the literature. \citet{sch98} considered the effect of frictional AML in recurrent nova outbursts, i.e. interaction of the expanding nova envelope with the donor, on the secular evolution of CVs. They showed that the strength of frictional AML is sensitive to the expansion
velocity $v_{\rm exp}$ of the envelope at the location of the donor, being stronger for smaller $v_{\rm exp}$. Indeed lower expansion velocities are expected for low-mass WD in CVs during nova outbursts than high-mass WDs \citep{livio91,yar05}. Alternatively, part of the slowly expanding ejected matter may form a CB disk or CE \citep{wil13,nel16}, and cause additional AML. Matter can also be transported into the circumbinary space via  a wind/outflow from the accretion disk or the secondary star as a natural consequence of the mass transfer.  Numerical simulations of the mass transfer process in CVs have demonstrated that as much as $ 50\%$ of the  transferred matter can escape from the WD's Roche lobe and end up in the circumbinary space around the CV \citep{b03,bk10}. Infrared observations of CVs by the {\em Spitzer Space Telescope} have revealed the presence of dust in many systems, indicating the possible origin of a CB disk \citep{h06,d07,h07,h09}. In this work we take into account the effect of  mass loss both from the WD and through a CB disk. We show that if around $(20-30)\%$ of the matter ejected during nova eruptions forms a CB disk, mass transfer in low-mass WDs are likely to be dynamically unstable.

Our results are largely consistent with \citet{nel16}, in which a CE phase is involved. While the frictional AML and CE models invoke discontinuous mass loss during nova eruptions, mass loss in the CB disk model can be both continuous and discontinuous. In the former case mass loss can be from the donor star rather the WD \citep{wil13}, and one would not expect substantial change in the orbital period after a nova outburst as in the frictional AML and CE models.
Finally we note that CB disks may also help explain the large spread for mass transfer rates in CVs for a given orbital period \citep{spr01}.

\acknowledgments We thank the referee for her/his valuable comments that helped to improve this paper. We are also grateful to Yong Shao for useful discussion. This work was supported by the Natural Science Foundation of China under grant numbers 11133001, 11333004 and U1331117, the National Key Research and Development Program of China (2016YFA0400803), and the Strategic Priority Research Program of CAS under grant No. XDB09000000.

\begin{figure}
\centering
\includegraphics[scale=0.43]{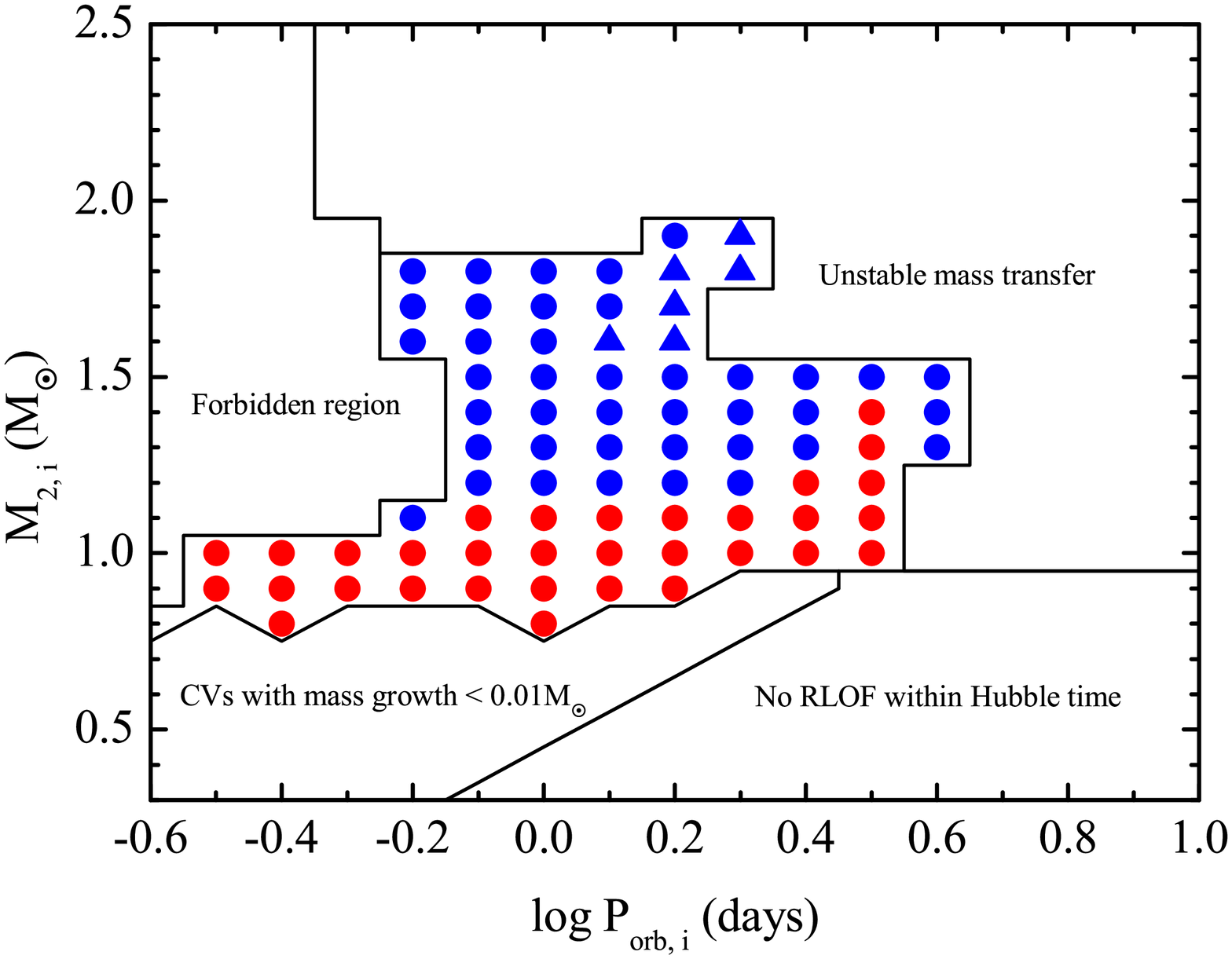}
\caption{Distribution of the initial orbital period and the
companion mass of the PCEBs for $M_{\rm {WD,~i}}$ = 0.6 $M_\odot$ with different outcome.
The red and blue circles represent the final WD mass of $0.6-0.8M_\odot$, and $0.8-1.0M_\odot$, respectively. The red and blue triangles denote the same final WD mass range respectively,  but the binaries cannot evolve into CVs.}
\label{fig:1}
\end{figure}

\clearpage

\begin{figure}
\centering
\includegraphics[scale=0.43]{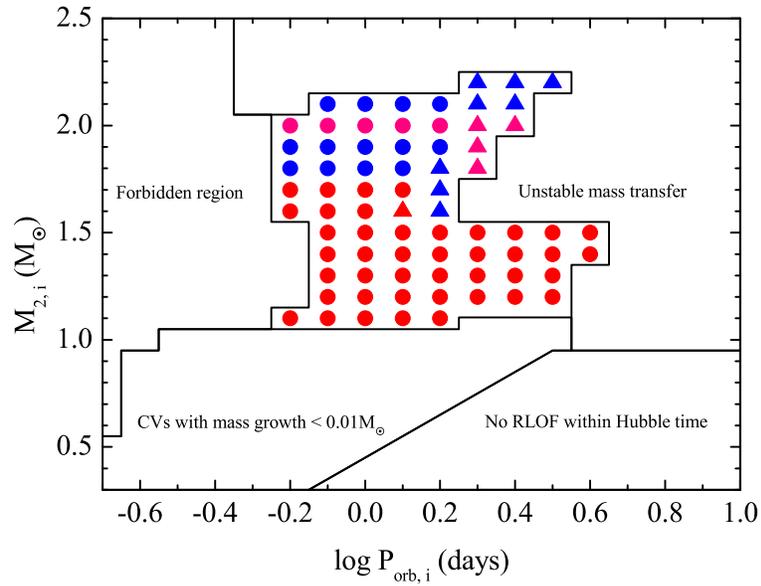}
\caption{Same as Fig. 1 but for $M_{\rm {WD,~i}}$ = 0.7 $M_\odot$.
The red, blue, and pink circles represent the final WD mass of $0.7-0.9M_\odot$, $0.9-1.1M_\odot$, and $1.1-1.2M_\odot$, respectively. The triangles have the similar meanings as in Fig.~1.}
\label{fig:2}
\end{figure}

\clearpage

\begin{figure}
\centering
\includegraphics[scale=0.30]{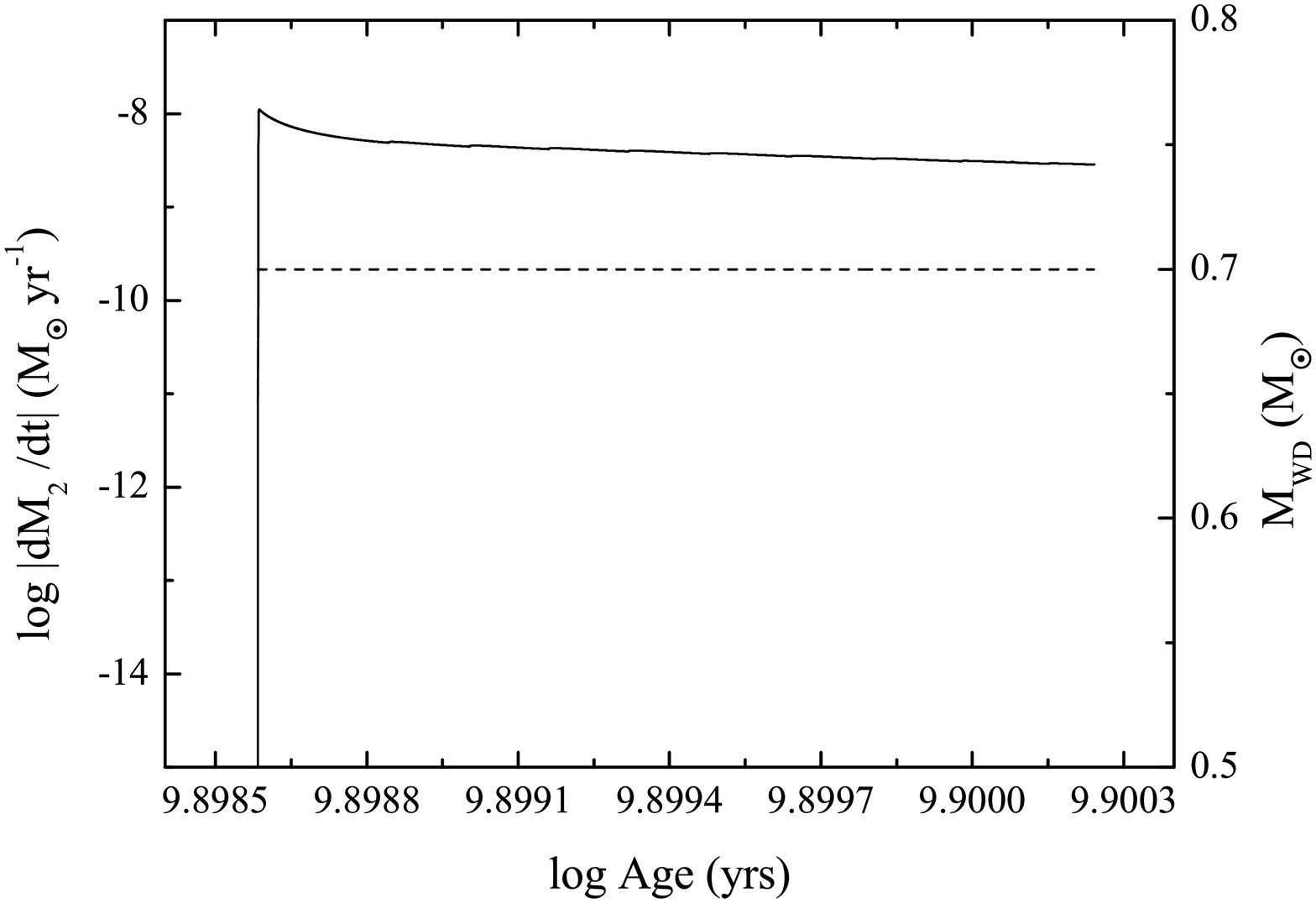}\includegraphics[scale=0.30]{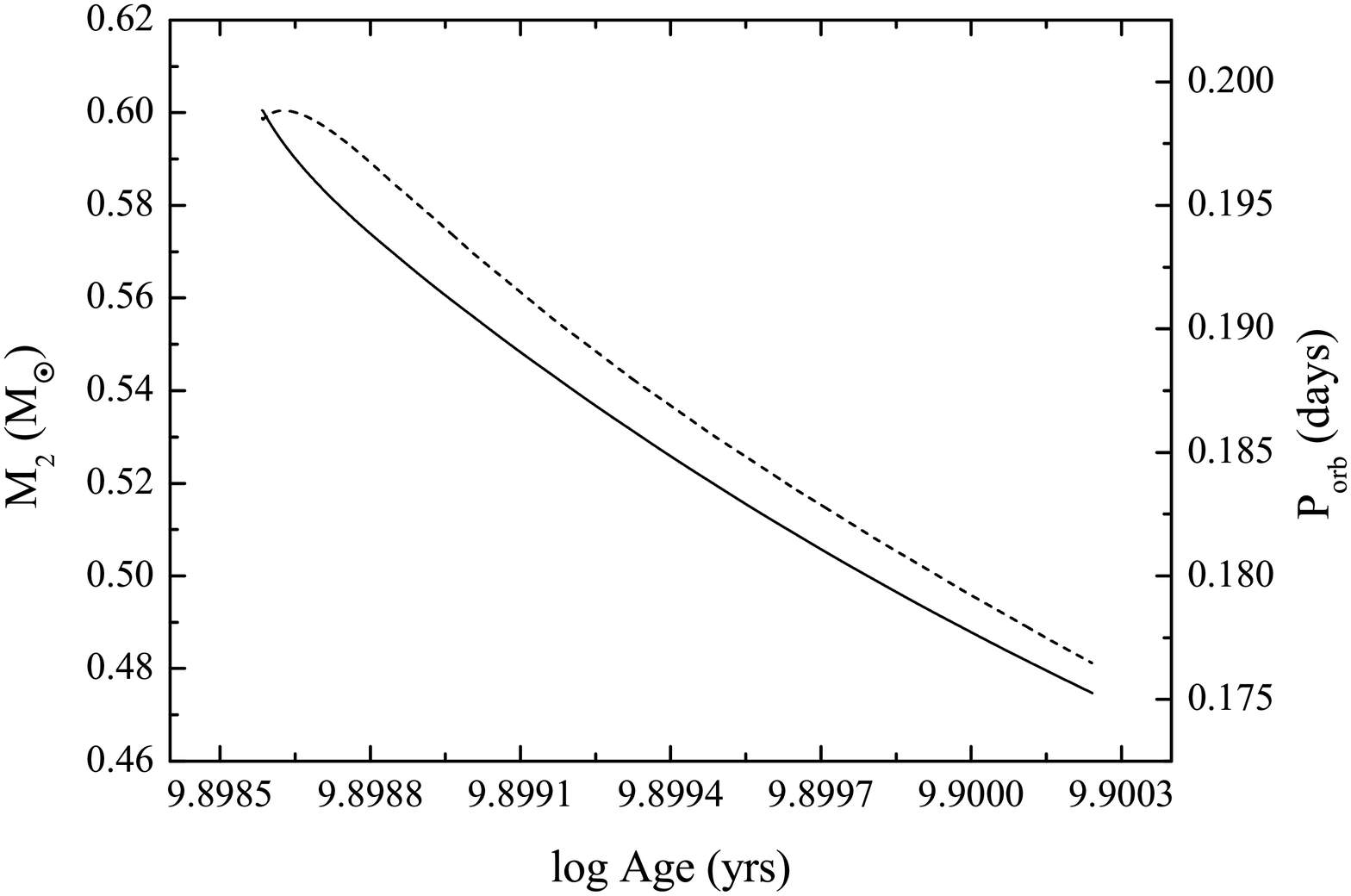}
\caption{Evolutionary track of a PCEB with $M_{\rm {2,~i}}$ = 0.6 $M_\odot$, $P_{\rm orb,~i}$ = 1.26 days and $M_{\rm {WD,~i}}$ = 0.7 $M_\odot$. The solid and dashed curves represent the evolution of mass transfer rate and the WD mass (left panel), the donor star mass and the orbital period (right panel), respectively}.
\label{fig:subfig}

\end{figure}

\clearpage

\begin{figure}
\centering
\includegraphics[scale=0.30]{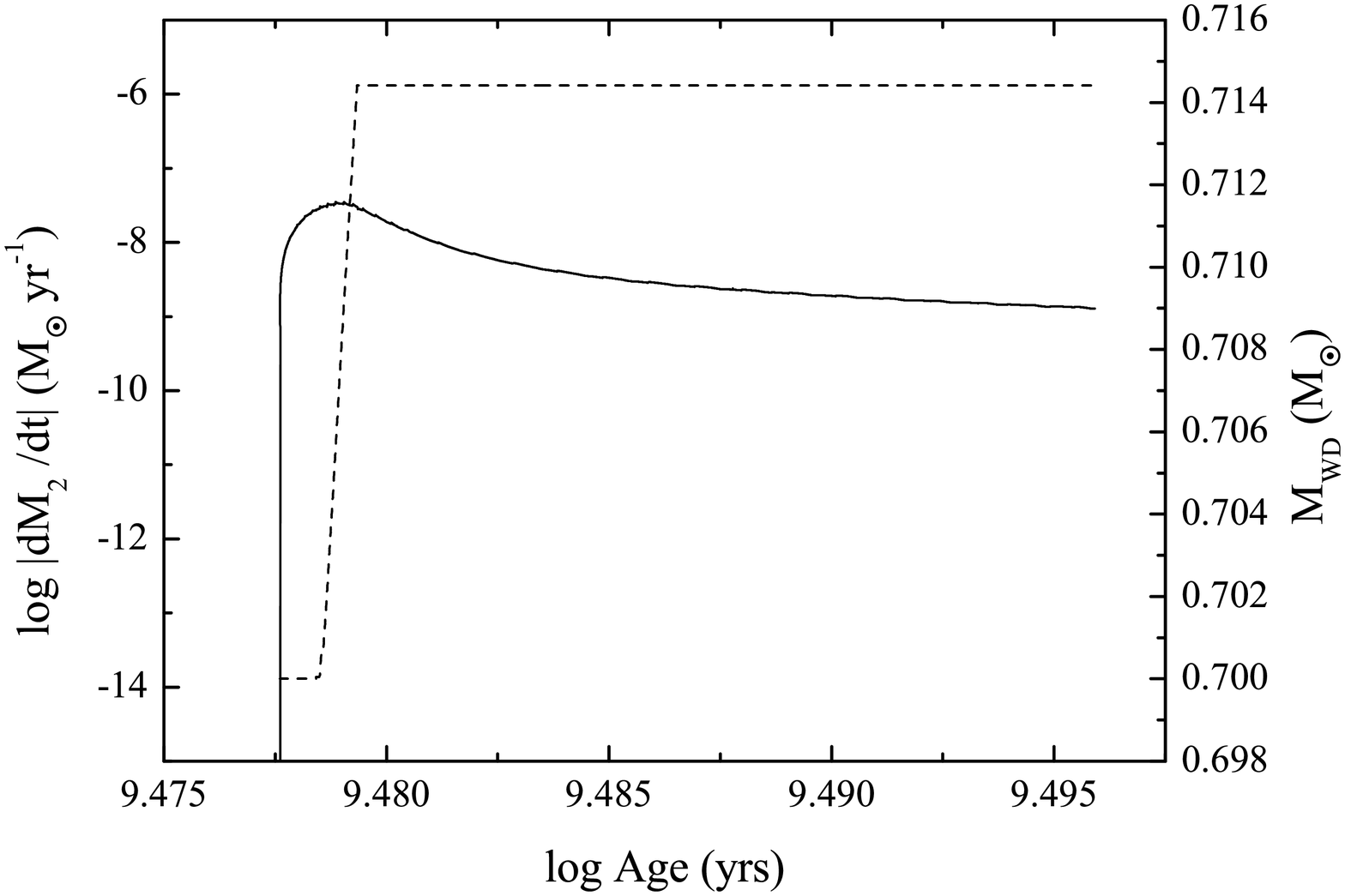}\includegraphics[scale=0.30]{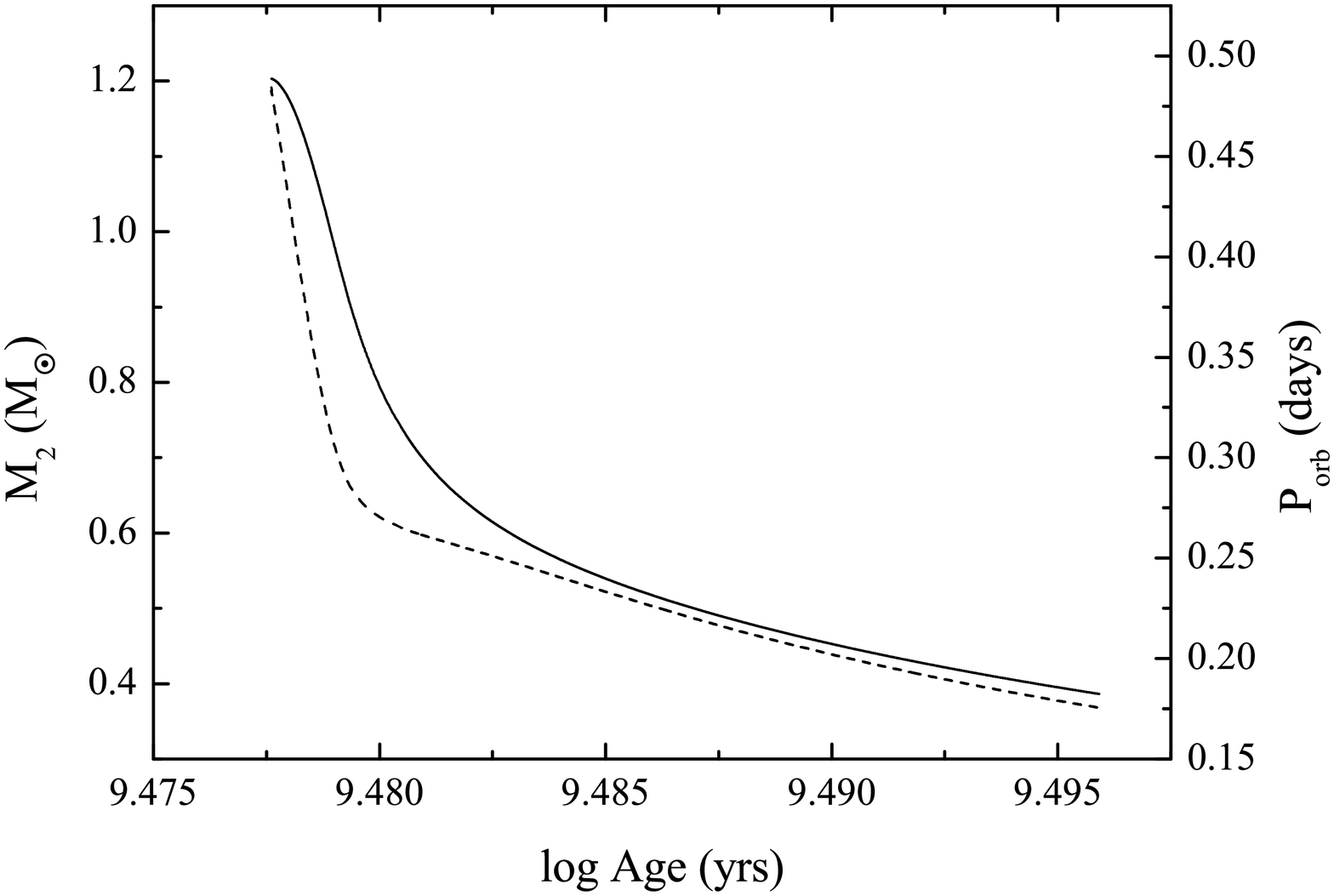}
\caption{Same as Fig. 3 but for $M_{\rm {2,~i}}$ = 1.2 $M_\odot$, $P_{\rm orb,~i}$ = 2.51 days and $M_{\rm {WD,~i}}$ = 0.7 $M_\odot$.}
\label{fig:subfig}

\end{figure}

\clearpage

\begin{figure}
\centering
\includegraphics[scale=0.30]{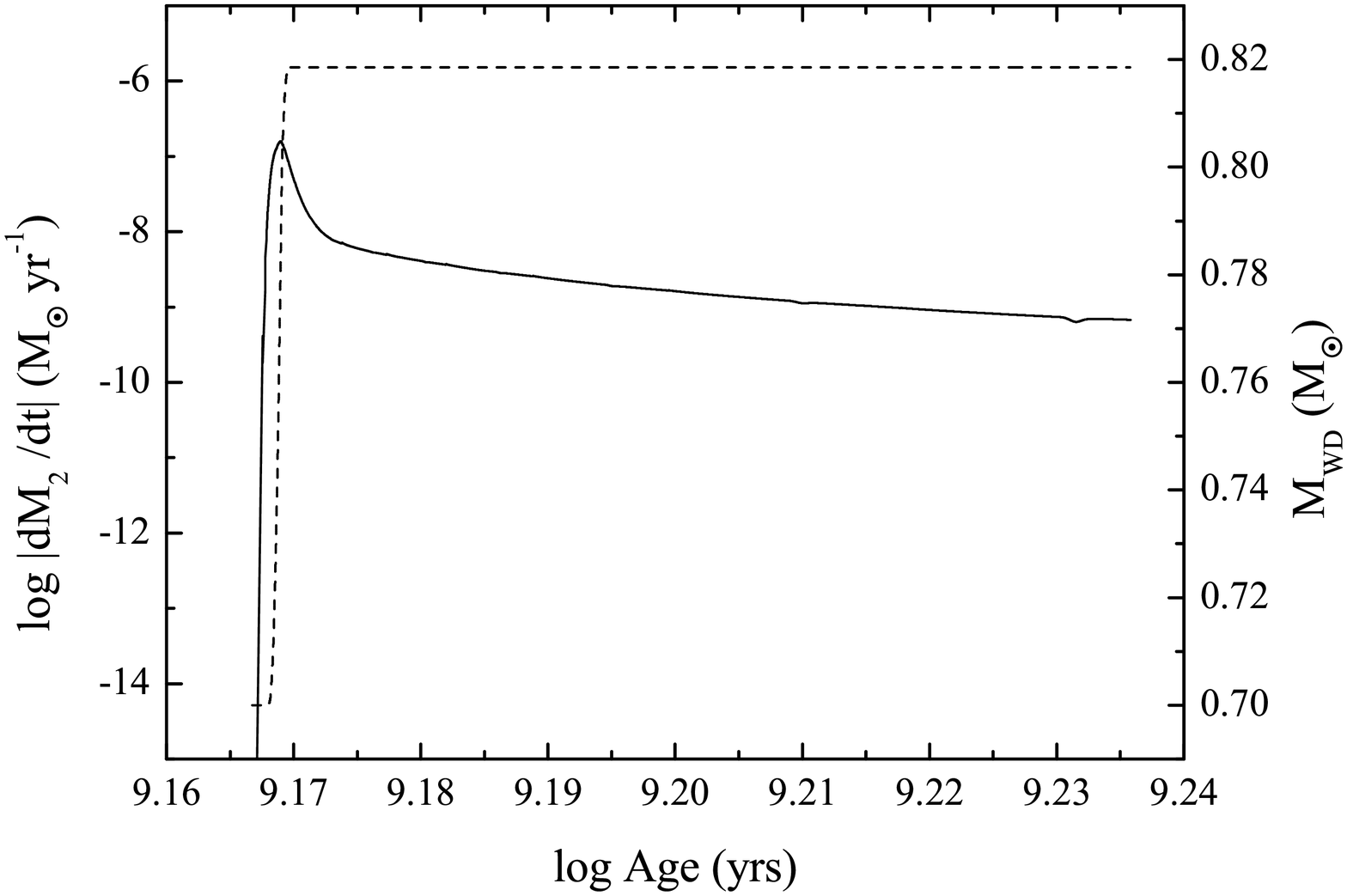}\includegraphics[scale=0.30]{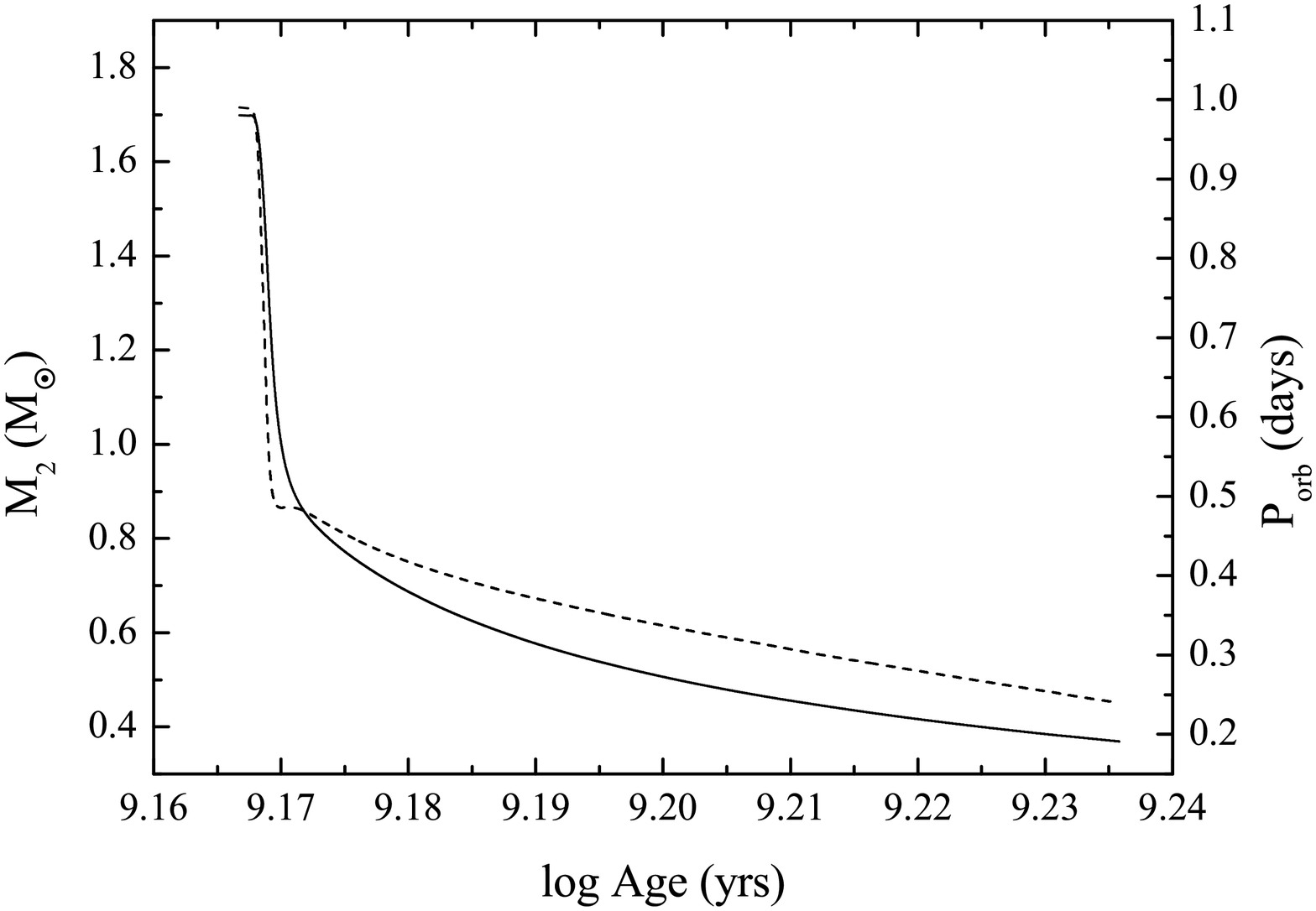}
\caption{Same as Fig. 3 but for $M_{\rm {2,~i}}$ = 1.7 $M_\odot$, $P_{\rm orb,~i}$ = 1.0 days and $M_{\rm {WD,~i}}$ = 0.7 $M_\odot$.}
\label{fig:subfig}

\end{figure}

\clearpage

\begin{figure}
\centering
\includegraphics[scale=0.30]{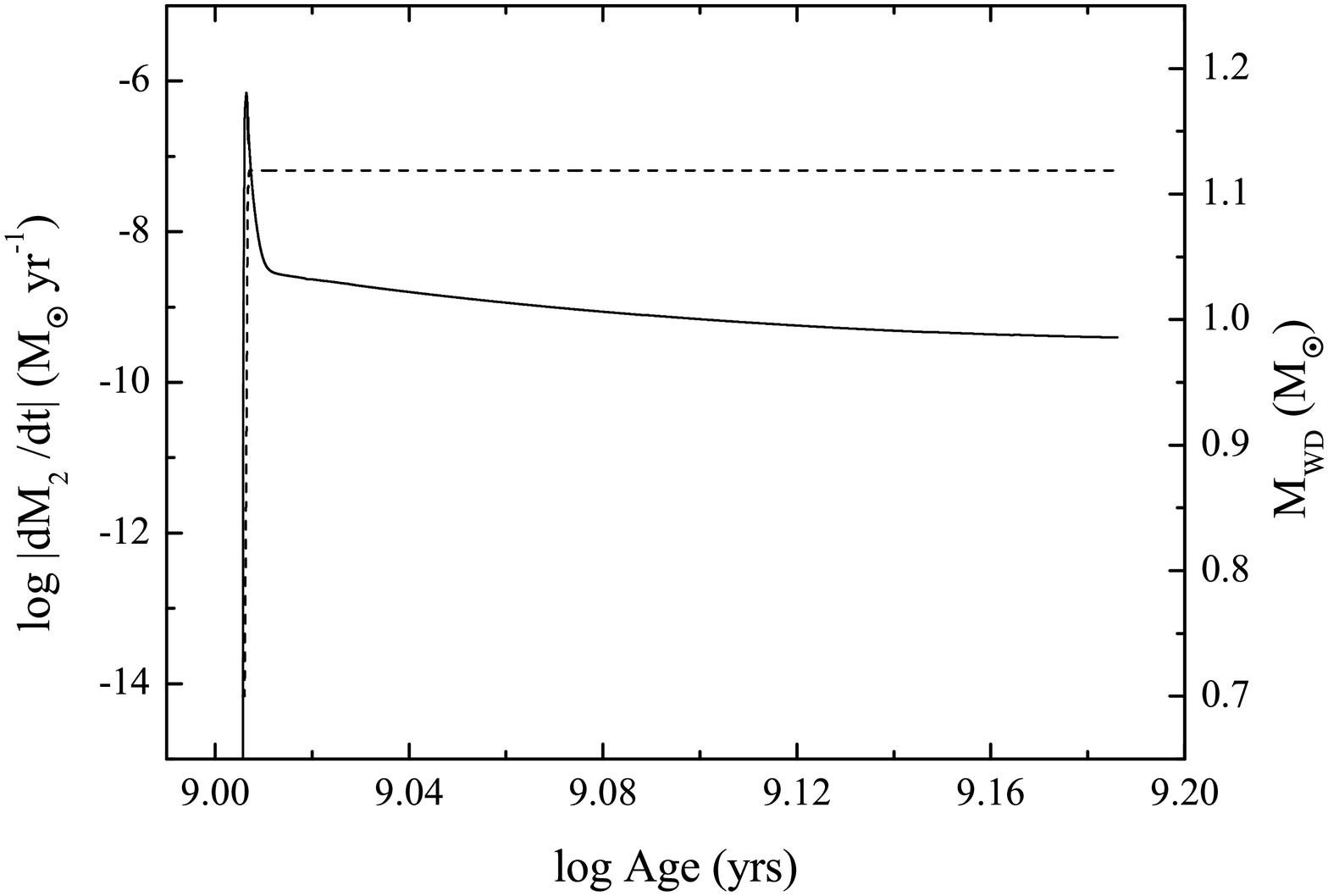}\includegraphics[scale=0.30]{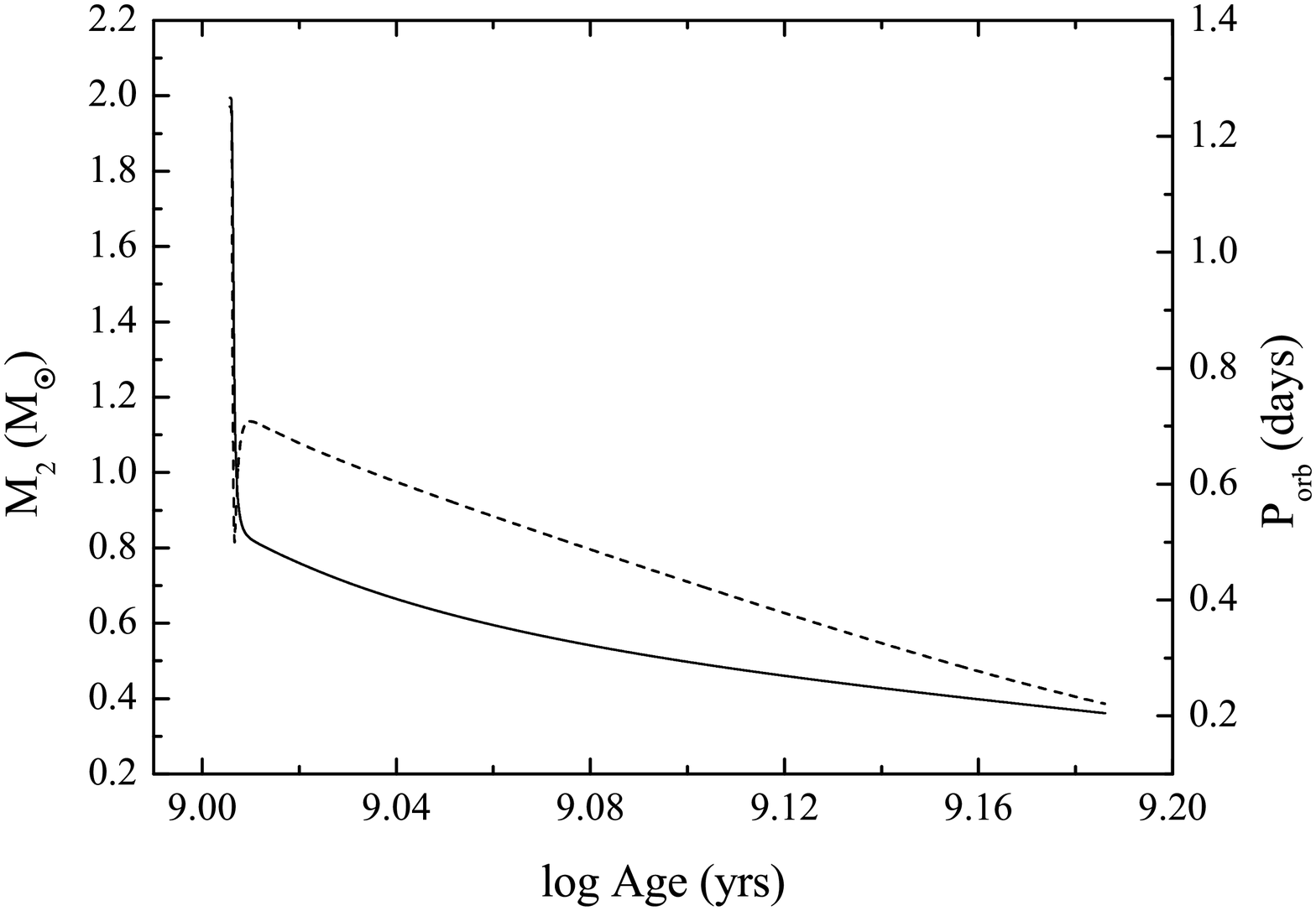}
\caption{Same as Fig. 3 but for $M_{\rm {2,~i}}$ = 2.0 $M_\odot$, $P_{\rm orb,~i}$ = 1.26 days and $M_{\rm {WD,~i}}$ = 0.7 $M_\odot$.}
\label{fig:subfig}

\end{figure}

\clearpage

\begin{figure}
\centering
\includegraphics[scale=0.30]{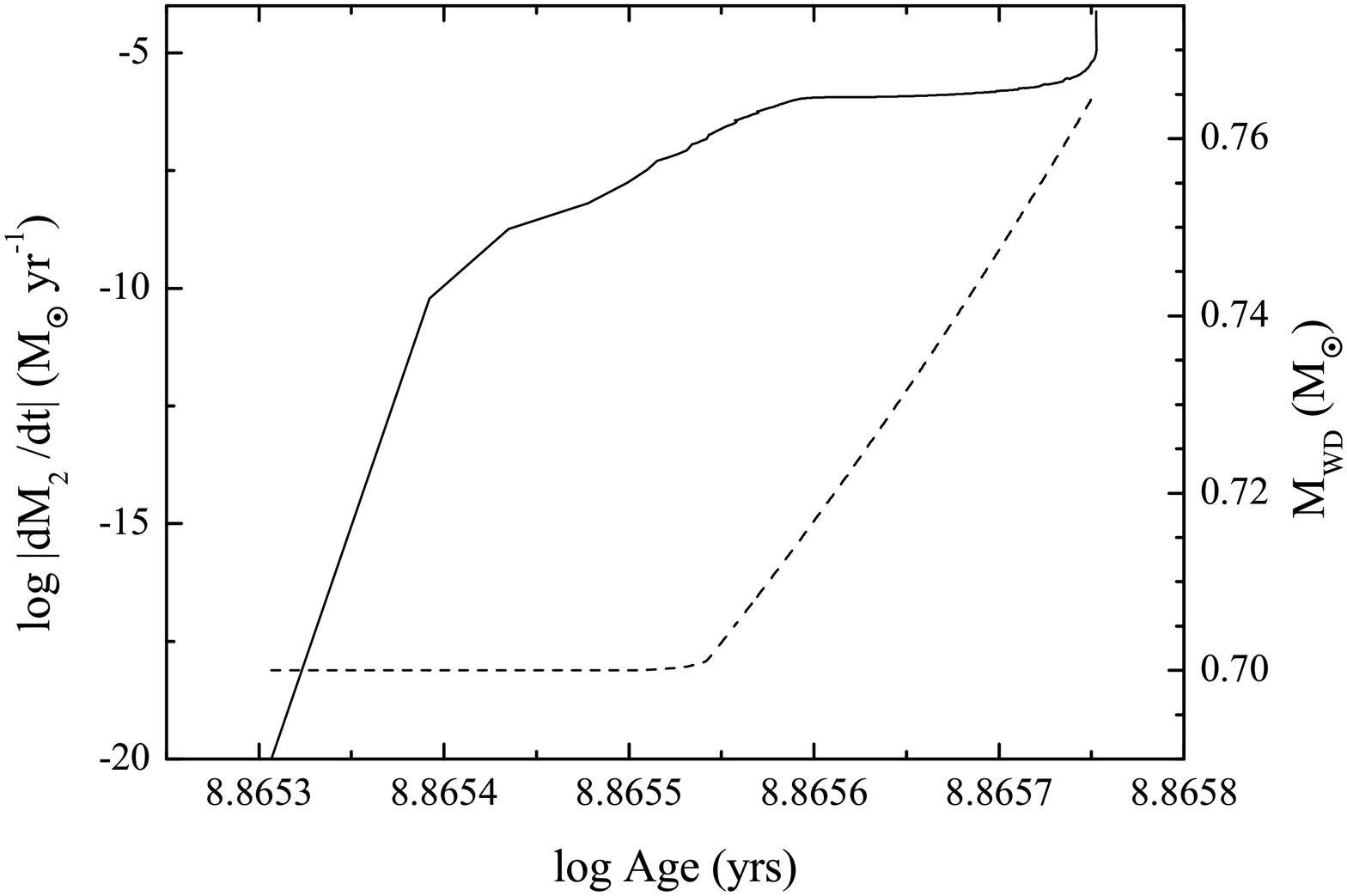}\includegraphics[scale=0.30]{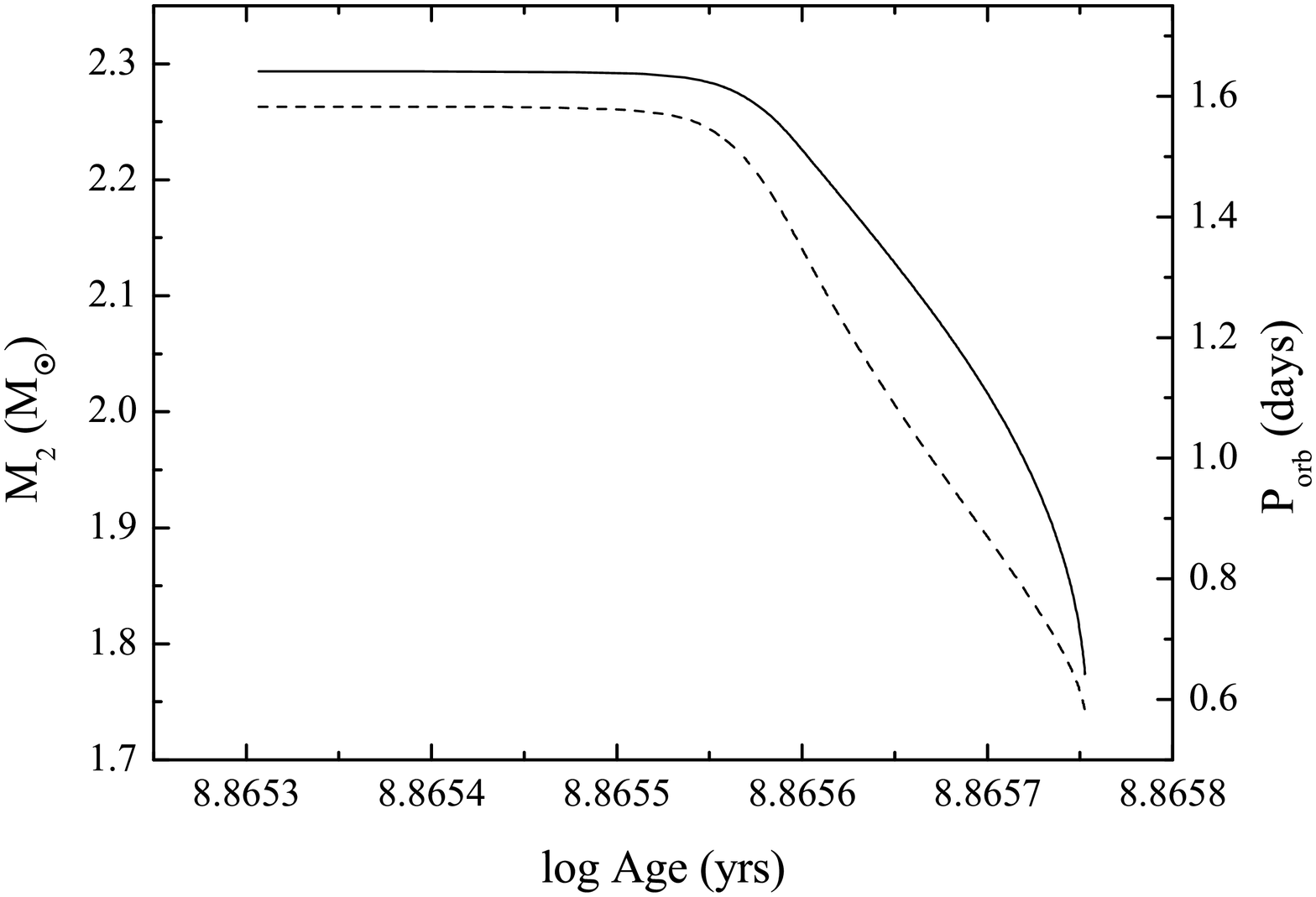}
\caption{Same as Fig. 3 but for $M_{\rm {2,~i}}$ = 2.3 $M_\odot$, $P_{\rm orb,~i}$ = 1.58 days and $M_{\rm {WD,~i}}$ = 0.7 $M_\odot$.}
\label{fig:subfig}

\end{figure}

\clearpage

\begin{figure}
\centering
\includegraphics[scale=0.43]{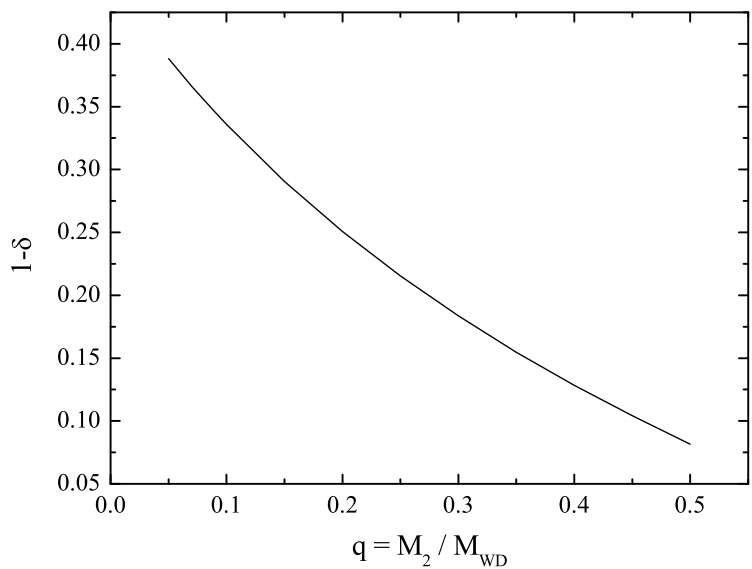}
\caption{Relation between $1-\delta$ and $q$ when both isotropic wind and CB disk are considered.}
\label{fig:8}

\end{figure}

\clearpage

\begin{figure}
\centering
\includegraphics[scale=0.43]{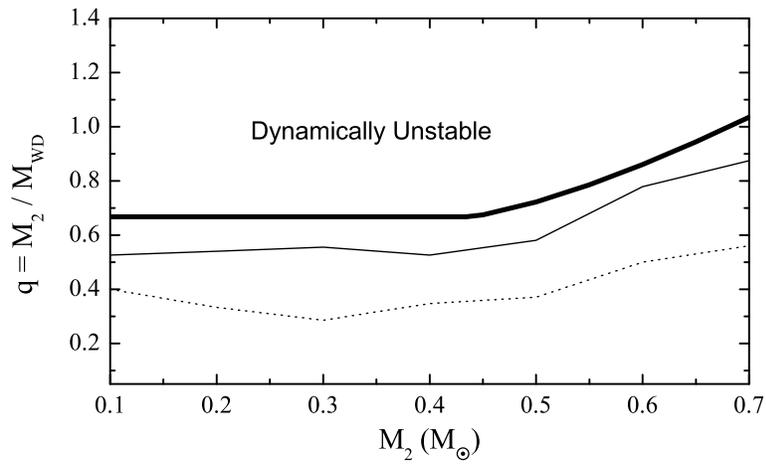}
\caption{The thin solid and dotted line demonstrate the boundaries between dynamically unstable and stable mass transfer in the $q$ versus $M_2$ diagram with $1-\delta$ = 0.2 and 0.3, respectively. The thick solid line shows the boundary for conventional mass transfer taken from \cite{pol96}.} \label{fig:subfig}

\end{figure}

\clearpage
\begin{figure}
\centering
\includegraphics[scale=0.43]{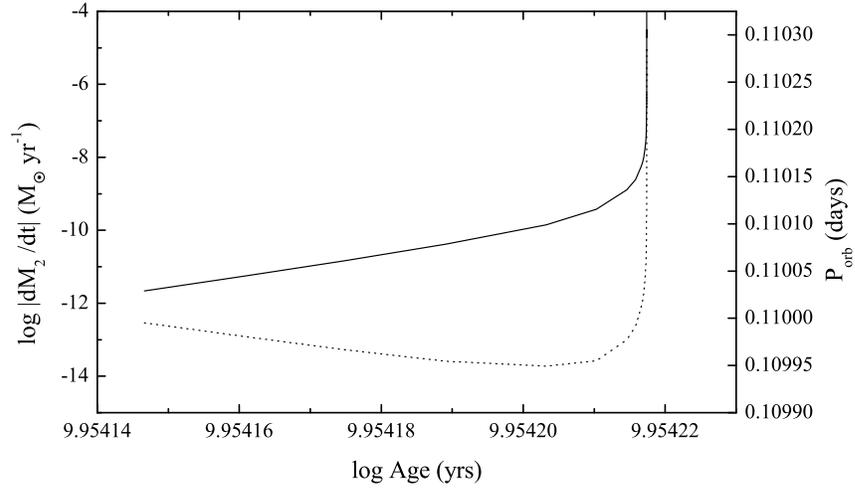}
\caption{Evolution of the mass transfer rate (solid line) and orbital period (dotted line) for a binary with $M_{\rm WD}=0.5\,M_\sun$, $M_2=0.3\,M_\sun$, $P_{\rm orb,~i}=0.28$ day, and $1-\delta=0.2$.} \label{fig:subfig}

\end{figure}
%
%
%

\label{lastpage}

\end{document}